# A Target-Free Harmonization Method for MRI

Minjun Kim[1]*, Dong Ju Mun[1]*, Hwihun Jeong[2], Hangyeol Park[1], Haechang Lee[1], Se Young Chun[1†], and Jongho Lee[1†]

[1]Department of Electrical and Computer Engineering, Seoul National University, Seoul, 08826, Republic of Korea

[2]Department of Psychiatry, Brigham and Women's Hospital, Harvard Medical School, Boston, MA, 02115, USA

## Correspondence

Jongho Lee, Ph.D

Department of Electrical and Computer Engineering, Seoul National University, Seoul 08826, Republic of Korea

E-mail: jonghoyi@snu.ac.kr

Se Young Chun, Ph.D

Department of Electrical and Computer Engineering, Seoul National University, Seoul 08826, Republic of Korea

E-mail: sychun@snu.ac.kr

* Co-first authors

† Co-corresponding authors

**Abstract**

In MRI, variations in scan parameters, sequence, or hardware can lead to discrepancies in image appearance, even for the same subject. These inconsistencies, known as domain shifts, can hinder image analysis and degrade the performance of deep learning models trained on data from specific target domains. MRI image harmonization aims to address these issues by aligning source domain images to the target domain images while preserving biological information such as anatomical structures. However, most existing harmonization approaches require access to both source and target domain data in training or test time. This dependence induces data sharing between institutions, raising concerns about patient privacy and substantially limiting the harmonization approaches that can be practically deployed in clinical settings. To overcome these limitations, we introduce TgtFreeHarmony, the harmonization framework tailored for *target-free* scenarios, eliminating the need for target domain data and any data sharing, enabling privacy-preserving harmonization directly within the source institution. Our approach estimates the target domain style by searching the manifold of MRI domain style constructed via a disentanglement-based generator using Bayesian optimization guided by the performance of a downstream task model, which is trained on target domain data. We evaluated our method on the brain tissue segmentation task across multiple institutes and demonstrated that it effectively harmonizes source images into target images, leading to improved downstream task performance. By enabling harmonization without any access to target-domain data, TgtFreeHarmony establishes a new direction of harmonization preserving data privacy that can be realistically deployed within clinical environments.

## 1. Introduction

Magnetic resonance imaging (MRI) is a prevalent medical imaging modality, serving a pivotal role in disease diagnosis, monitoring, and treatment planning. Recent advances in deep learning have significantly enhanced automated MRI image analysis, facilitating more accurate and robust approaches. However, one of the major obstacles for deploying these models in a real-world clinical setting is the domain shift problem: MRI data exhibits substantial variations across different vendors, scanners, and scan parameters even when imaging the same subject (Cai et al., 2021). Consequently, a model trained on one domain (referred to as a target domain) often demonstrates significantly degraded performance when applied to data from other domains (referred to as source domains).

Efforts to reduce domain gaps have increasingly focused on training generalizable models using large multi-center datasets that span diverse scanners and acquisition conditions (Li et al., 2023; Ma et al., 2024; Moor et al., 2023). However, they require extensive multi-institutional data aggregation and cross-institutional data sharing, which may raise patient data privacy concerns. Consequently, alternative strategies have been explored to address domain shift. One such approach is traditional transfer learning through fine-tuning that utilizes paired images and labels from the source domain to adapt pre-trained models (Tajbakhsh et al., 2016). However, fine-tuning becomes unreliable when only a small number of labeled data are available, as models tend to overfit and fail to generalize across domains (Kornblith et al., 2019). Domain adaptation techniques are another approach, aligning source and target feature distributions (Ben-David et al., 2006; Long et al., 2015), but these methods are primarily operated in feature space; they may neglect to preserve clinically critical information, such as anatomical structure, potentially leading to unreliable downstream clinical interpretation, such as inaccurate tissue segmentation or volumetric measurements (Cui et al., 2021; Wang and Deng, 2018).

Image harmonization methods offer an alternative strategy that aligns images from source domains to match a target domain, reducing domain-specific biases while preserving biological information. Preprocessing methods such as histogram matching (Nyúl et al., 2000) and statistical normalization (Shinohara et al., 2014) can be considered as conventional harmonization methods. In deep learning, DeepHarmony (Dewey et al., 2019) employs supervised learning from traveling subjects scanned across domains. Unsupervised methods (Guan et al., 2021; Liu et al., 2021; Modanwal et al., 2020; Roca et al., 2025; Zuo et al., 2021b) eliminate this requirement by learning from unpaired data in each domain. Both supervised and unsupervised approaches, however, require access to data from source and target domains, incurring substantial data collection costs and inter-institutional data sharing, which poses risks to clinical data privacy. More recently, source-free harmonization methods (Beizaee et al., 2025; Jeong et al., 2023) have been introduced, which are trained exclusively on target domain data and adapt source inputs to the learned target distribution at test time. Although these methods eliminate the need for source data during training, they are inherently developed from the perspective of downstream model providers who possess the target domain data. As a result, data-holding institutions wishing to use an external downstream task model must either share their source domain data with the provider or receive a harmonization model already trained on the target domain, which assumes prior access to target domain data (Table 1). Even in the latter case, privacy risks remain, as model parameters can leak sensitive information through model inversion attacks (Fredrikson et al., 2015; Zhang et al., 2020). Other approaches, which train harmonization models by disentangling domain-invariant anatomical content from domain-dependent style, defined as scanner- or site-specific image appearance factors such as contrast (Cackowski et al., 2023; Dewey et al., 2020; Scholz et al., 2025), have been proposed, enabling training with only source domain data and adaptation at test time via style transfer from target domain images. However, because they still require target domain data as a reference during inference, they are not fully target-free. As a result, all existing methods inherently require either data sharing or the use of models trained on target domain data, which introduces data-privacy challenges and limits their adoption in clinical settings.

To address this challenge, we introduce Target Free Harmony or TgtFreeHarmony, the harmonization framework explicitly designed for a target-free scenario, where direct access to the target domain data is unavailable. Instead of relying on target data, we assume that the performance of a downstream task model, which is trained on the target domain data, is closely linked to the magnitude of domain shift between target and source domains (i.e., a larger domain shift means poorer performance in the source domain data). Based on this assumption, we leverage this performance-domain shift connection as indirect information to guide harmonization. The harmonization process is therefore formulated to adjust image appearance in a direction that improves downstream task performance, progressively reducing the domain shift and increasing similarity to the target domain. Accordingly, TgtFreeHarmony performs harmonization using downstream task performance computed from a small amount of labeled source domain data. Our key contributions are as follows:

• We present TgtFreeHarmony, which is the harmonization method designed for the target-free setting, enabling harmonization to be developed and deployed directly by data-holding institutions.

• TgtFreeHarmony introduces a novel formulation that links downstream task performance to domain shift for target-free harmonization.

• TgtFreeHarmony is built upon utilizing a generative framework that constructs an MRI style manifold and searches it to estimate the target style, guided by downstream task performance.

• Extensive experiments on two large-scale open-source datasets (OASIS-3 and SRPBS; three vendors and six scanners) demonstrate that the proposed method enables successful harmonization for a brain tissue segmentation downstream task.

**Table 1.** Categorization of domain shift reduction methods according to the required data (image $x$, label $y$) from source ($^s$) and target ($^t$) domains, and whether data or model sharing is required. The source domain refers to data unseen during downstream model training, while the target corresponds to the domain in which the downstream model was originally trained.

| Setting | Data requirements | | Data or model sharing |
|---|---|---|---|
| | Source data | Target data | |
| Supervised harmonization | $x^s$ | $x^t$ | Required |
| Unsupervised harmonization | $x^s$ | $x^t$ | Required |
| Source-free harmonization | $x^s$ (in test time) | $x^t$ | Required |
| Target-free harmonization (ours) | $x^s, y^s$ | Not required | Not required |

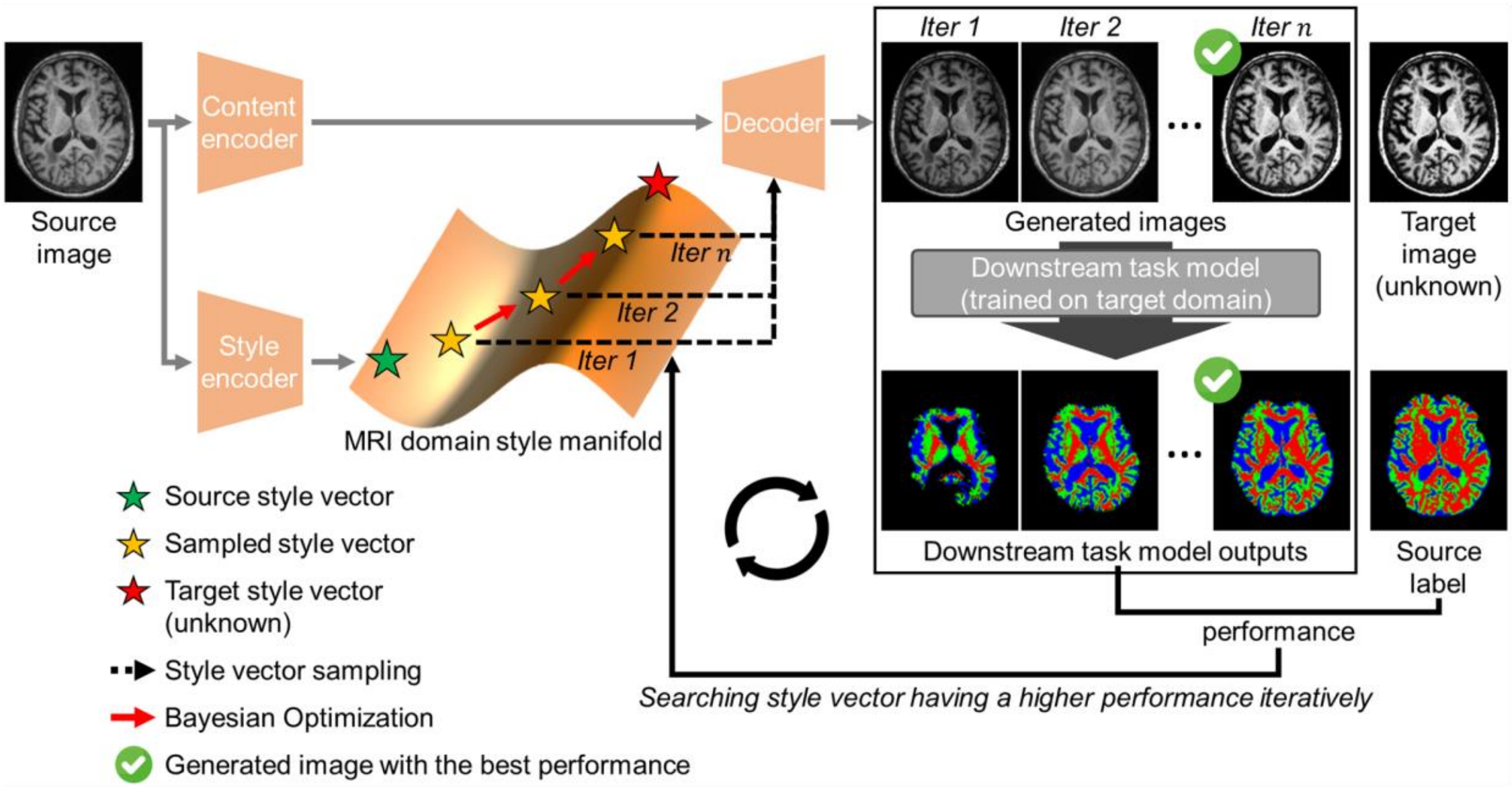

**Figure 1.** Overview of the proposed target-free harmonization method. The disentanglement-based generator comprises content/style encoders and a decoder. The content encoder extracts domain-invariant information, such as anatomical structure, while the style encoder captures scanner- or site-dependent appearance factors, including contrast, and organizes them into an MRI style manifold. By sampling style vectors from this manifold and combining them with the content feature in the decoder, the generator synthesizes images that preserve structure while altering image style. This manifold serves as the basis for harmonization, where style vectors are searched iteratively in a direction that improves downstream task performance to identify the representation that best approximates the unknown target domain. The image generated with the highest-performing style vector is selected as the final harmonized output.

## 2. Methods

Fig. 1 illustrates the overall pipeline of TgtFreeHarmony. The framework employs a disentanglement-based generator comprising a content encoder, a style encoder, and a decoder. The content encoder extracts a domain-invariant feature (i.e., content feature), such as anatomical structure, while the style encoder captures a domain-variant feature (i.e., style feature or style vector), including contrast, blurriness, noisiness, etc., and organizes them into an MRI domain style manifold. The decoder combines the content feature and a sampled style vector to synthesize an image of a different style (see Section 2.1). Then, the image is evaluated using a downstream task model, and its performance is computed with respect to the source domain label. This performance serves as a reward to search the MRI domain style manifold for a style vector that best approximates the target domain. This search is performed iteratively, and the image generated with a style vector yielding the best downstream task performance is selected as the final harmonized output (see Section 2.2).

### 2.1. MRI Domain Style Manifold Construction using Disentanglement-based Generator

Our generator is built upon the content-style disentanglement framework (Huang et al., 2018), consisting of a content encoder $E_c$, a style encoder $E_s$, and a decoder $D$. For an input MRI image $x$, the encoders extract a content feature $z_c \in R^{h \times w}$ and a style vector $z_s \in R^d$, given by $z_c = E_c(x)$ and $z_s = E_s(x)$, where $h$ and $w$ denote the height and width of the content feature map and $d$ represents the dimensionality of the style vector. Following prior harmonization studies (Cackowski et al., 2023; Dewey et al., 2020; Liu and Yap, 2024; Zuo et al., 2021a; Zuo et al., 2021b), we adopt the definition of domain-invariant content as the underlying anatomical structures in MRI images, while domain-variant style as image appearance factors that contribute to inter-domain variation, such as contrast, blurriness, and noisiness (Kushol et al., 2023). The decoder reconstructs the input image by combining the content and style features:

$$x = D(z_c, z_s) = D\big(E_c(x), E_s(x)\big). \quad (1)$$

To construct an MRI domain style manifold, the latent space of the style encoder is mapped to a Gaussian distribution. Then, a style vector sampled from this distribution is combined with the content feature in the decoder to generate an MRI image with a different domain style:

$$x' = G(z'_s ; z_c) = D(E_c(x), z'_s) , \quad z'_s \sim N(0, I), \tag{2}$$

where $G(\cdot)$ is the generator and $x'$ is a generated image, whose anatomical structure is preserved with the input image $x$ while its style is determined by the sampled style vector $z'_s$.

For the training of the generator, supervised learning is performed using pairs of MRI images $x$ and their perturbed counterparts $\tilde{x}$, following prior studies that employed contrast-based perturbations for disentanglement learning (Cackowski et al., 2023; Scholz et al., 2025). While these works focused primarily on contrast adjustment, we extended the perturbation strategy to include random combinations of intensity scaling, gamma correction, Gaussian blurring, and Gaussian noise addition. Specifically, after normalizing the image intensity to the range [0, 1], intensity perturbation is applied by scaling the values with a factor sampled between 0.5 and 1.5, followed by the addition of an offset randomly sampled between −0.25 and 0.25, and subsequent gamma correction with gamma values between 0.7 and 1.3. Gaussian blur is then applied with a standard deviation sampled between 0 and 1. Finally, zero-mean Gaussian noise with a standard deviation between 0 and 0.03 is added. During training, the original and perturbed images are intentionally swapped as inputs, encouraging the generator to learn not only the applied perturbations but also their inverse transformations.

The generator is trained using four loss functions following Huang et al. (2018): disentanglement loss ($L_{disent}$), reconstruction loss ($L_{recon}$), adversarial loss ($L_{adv}$), and KL-divergence loss ($L_{KL}$). The disentanglement loss enforces consistency between the content features and style vectors extracted from the input and reconstructed images:

$$L_{disent} = \|z_c - \tilde{z_c}\|_1 + \left\|z_c - E_c\left(D(z_c, \tilde{z_s})\right)\right\|_1 + \left\|z_s - E_s\left(D(\tilde{z_c}, z_s)\right)\right\|_1, \quad (3)$$

where $\tilde{z_c}$ and $\tilde{z_s}$ denote the content feature and style vector extracted from the perturbed image, respectively. The reconstruction loss is designed to ensure faithful image reconstruction while preserving the structural content of the input image:

$$L_{recon} = \|x - D(z_c, z_s)\|_1 + \|x - D(\tilde{z_c}, z_s)\|_1 + \left\|x - D\left(E_c\left(D(\tilde{z_c}, z_s)\right), z_s\right)\right\|_1. \quad (4)$$

To promote realistic image synthesis and regularize the style manifold to follow a Gaussian distribution, adversarial and KL-divergence losses are incorporated:

$$L_{adv} = -\left(log\ Dis(x)\ + log\left[1 - Dis\left(D(z_c, \tilde{z_s})\right)\right]\ + log\ Dis(\tilde{x})\ + log\left[1 - Dis\left(D(\tilde{z_s}, z_s)\right)\right]\right), \quad (5)$$

$$L_{KL} = D_{KL}\left(z_s \| N(0, I)\right) + D_{KL}\left(E_s\left(D(\tilde{z_c}, z_s)\right) \| N(0, I)\right), \quad (6)$$

where $Dis(\cdot)$ denotes the discriminator. As described earlier, $x$ and $\tilde{x}$ are used interchangeably as inputs across the loss terms to promote diverse style coverage and symmetric disentanglement learning. The overall training objective is the weighted sum of the four losses, with weights of 1, 1, 0.1, and 0.01 for $L_{disent}$, $L_{recon}$, $L_{adv}$, and $L_{KL}$, respectively.

This generator serves as the foundation of our harmonization framework, enabling searching for a wide range of target domain styles.

## 2.2 Estimating an Unknown Target Domain Style over the Generator's Style Manifold

To estimate the best approximation of the target domain style within the generator's style manifold, the performance of the downstream task model is used as guidance. The target style estimation problem is formulated as the following optimization objective:

$$z^* = \arg max_{z_s'} \varphi_{f_{task}}(\ G(z_s';\ z_c), y), \quad (7)$$

where $\varphi_{f_{task}}$ denotes the downstream task performance, $z^*$ represents the estimated target style vector, and $y$ is the source domain label. This objective aims to identify the style vector that maximizes downstream task performance, under the assumption that such a style most closely approximates the unknown target domain. Examples of downstream task performance include metrics such as the Dice score for segmentation tasks.

Searching for the optimal style vector based on Eq. (7) requires repeated evaluation of the downstream model. To be efficient, Bayesian optimization is adopted (Brochu et al., 2010; Frazier, 2018). A Gaussian process (GP) model (Williams and Rasmussen, 2006) is employed for the search. The GP model is first initialized using a dataset containing sampled style vectors and their corresponding downstream task performance. Based on this model, candidate style vectors are iteratively selected according to the objective function in Eq. (7). Each candidate is then evaluated using the downstream task model, and the resulting style vector-performance pair is added to the dataset to update the GP model. This process is repeated until convergence. After optimization, the resulting style vector is taken as the estimated target domain style, and the final harmonized image is generated.

## 2.3. Experiments

### 2.3.1. Experimental setting

For experiments, brain tissue segmentation was selected as the downstream task, where white matter, gray matter, and cerebrospinal fluid (CSF) were segmented. This task provides a direct measure of not only how well the proposed method aligns images across domains, but also whether anatomical structures are preserved under the disentanglement-based framework.

**Datasets.** Two open-source datasets, OASIS-3 (LaMontagne et al., 2019) and SRPBS (Tanaka et al., 2021), which include traveling-subject data acquired across multiple scanners, were used for the experiments. We utilized T1-weighted images from four Siemens scanners (Tim Trio, Sonata, Vision, and Vida) in OASIS-3 and three vendors (Siemens Tim Trio, Phillips Achieva, and GE Signa HDxt) in SRPBS. For the Siemens Vida scanner, only images acquired using sagittal MPRAGE were included, excluding images of other T1-weighted sequences. Additional dataset information is summarized in Table 2. All images were resampled to a uniform voxel size ($1 \times 1 \times 1$ mm$^3$) and normalized using slice-wise 1% percentile normalization. Ground-truth brain tissue labels were obtained using the FSL-FAST segmentation tool (Jenkinson et al., 2002). Labels for subjects with severe Alzheimer's disease were generated using SynthSeg (Billot et al., 2023), as FSL-FAST often fails to produce reliable segmentation in these cases.

Generator training used only Siemens Vision and Vida images from OASIS-3, with 150 subjects sampled from each scanner and split into training and validation sets at a 9:1 ratio. Images from the remaining scanners were excluded during generator training and later used as unseen domains to evaluate generalization.

Siemens Tim Trio, Philips Achieva, and GE Signa HDxt were designated as target domains for training the downstream segmentation models. The remaining datasets were treated as source domains, and only one subject from each source domain was used for target style estimation.

Traveling-subject data were reserved exclusively for evaluation. These subjects, which did not overlap with any training datasets, included multiple scanner pairs from OASIS-3 (Siemens Sonata–Tim Trio pairs: 21 subjects; Siemens Vision–Tim Trio pairs: 66 subjects;

Siemens Vida–Tim Trio pairs: 31 subjects) and SRPBS (Philips Achieva–Siemens Tim Trio pairs: 6 subjects; GE Signa HDxt–Siemens Tim Trio pairs: 6 subjects).

**Table 2.** Data descriptions of the dataset from OASIS-3 and SRPBS.

| Dataset | OASIS-3 | | | | SRPBS | | |
|---|---|---|---|---|---|---|---|
| Manufacturer | Siemens | Siemens | Siemens | Siemens | Siemens | Phillips | GE |
| Scanner | Tim Trio | Sonata | Vision | Magnetom Vida | Tim Trio | Achieva | Signa HDxt |
| Magnetic field strength (T) | 3 | 1.5 | 1.5 | 3 | 3 | 3 | 3 |
| Matrix size | 176 × 256 × 256 | 160 × 256 × 256 | 128 × 256 × 256 | 176 × 240 × 256 | 240 × 256 × 256 | 170 × 256 × 256 | 180 × 256 × 256 |
| TR/TI (ms) | 2400/100 | 1900/110 | 9700/unknown | 2300/unknown | 2300/900 | unknown/503 | 6788/450 |
| TE (ms) | 3.2 | 3.9 | 4.0 | 3.0 | 3.0 | 3.3 | 1.9 |
| Flip angle(°) | 8 | 15 | 10 | 9 | 9 | 10 | 20 |

**Implementation details.** The generator's content encoder consists of three convolutional layers with instance normalization (Ulyanov et al., 2016), followed by four residual blocks (He et al., 2016). The style encoder includes three convolutional layers, a global average pooling layer, and a fully connected layer that produces a 32-dimensional style vector, while the decoder consists of two upsampling–convolution blocks followed by a final convolutional layer. For style injection, the decoder incorporates residual blocks equipped with adaptive instance normalization (AdaIN) (Huang and Belongie, 2017). During generation, style vectors are sampled from a Gaussian distribution with zero mean and unit variance. The generator was trained for 200 epochs using the Adam optimizer (Kingma, 2014) with a batch-size of 16 and an initial learning-rate of $1 \times 10^{-4}$, which was decayed every $1 \times 10^{5}$ iterations using a StepLR scheduler. The final model was selected as the one with the lowest validation loss.

For Bayesian optimization, we used a Gaussian process model from GPyTorch (Gardner et al., 2018) with an automatic relevance determination and a radial basis function kernel (Williams and Rasmussen, 2006), tailored to the dimension of the style vector. GP-UCB (Srinivas et al., 2009) was utilized as the acquisition function with a confidence parameter $\beta = 0.1$. The Gaussian process model was initially trained with 100 random samples for 50 iterations using the Adam optimizer with a learning rate of 0.1. The optimization was performed for 100 iterations, using the Dice score as the guiding metric.

The segmentation model was implemented as a 2D U-Net (Ronneberger et al., 2015) with architectural details provided in Section S1. The model was trained using the cross-entropy loss, the Adam optimizer, and a learning rate of $1 \times 10^{-4}$ for 100 epochs, with a batch size of 32.

All experiments were conducted in PyTorch on a single NVIDIA L40S GPU.

### 2.3.2. Evaluation of the Disentanglement-Based Generator.

We evaluated the generator's disentanglement capability and its ability to produce diverse MRI styles.

Disentanglement capability was assessed through style transfer experiments using synthesized images generated by combining content and style features extracted from two different images. These images were examined to determine whether the model preserves the structural information of the content image while transferring the appearance of the style image. The evaluation was conducted using both paired traveling-subject and unpaired subjects.

To investigate if the generator's style manifold exhibits controllable variation, style vectors derived from two MRI images were linearly interpolated within the range [0, 1] at intervals of 0.33 and extrapolated beyond this range to [-0.5, 1.5] with a step size of 0.25. The resulting images were qualitatively examined to assess whether their appearance changed smoothly, as commonly evaluated in disentanglement studies (Chen et al., 2016). This analysis was performed using both traveling-subjects from the open-source datasets and synthetic image pairs generated with the perturbation scheme described in Section 2.1, which introduced variations in brightness, contrast, blurriness, and noisiness (see Section S2).

The diversity of the generated styles was evaluated using t-SNE. Generated images were synthesized by combining content features extracted from the evaluation datasets described in Section 2.3.1 with randomly sampled style vectors. Both the generated images and the evaluation dataset images were then passed through the style encoder to obtain style features, and t-SNE was applied to visualize their distributions using a perplexity of 30, a learning rate of 200, and PCA initialization.

### 2.3.3. Evaluation of Harmonization

The peak signal-to-noise (PSNR) and structural similarity (SSIM) were utilized to evaluate the similarities between the output and their corresponding target image. To assess the segmentation performance, intersection over union (IoU) and Dice score were employed. The baselines for comparison included conventional histogram matching (Nyúl et al., 2000), DeepHarmony (Dewey et al., 2019), style transfer (Liu et al., 2021), BlindHarmony (Jeong et al., 2023), and Harmonizing flows (Beizaee et al., 2025). All learning-based methods were trained using the same dataset described in Section 2.3.1, with one subject per source domain, to ensure a fair comparison. For methods that require access to both source and target domains, such as supervised and unsupervised approaches, one subject from each target domain was additionally used for training.

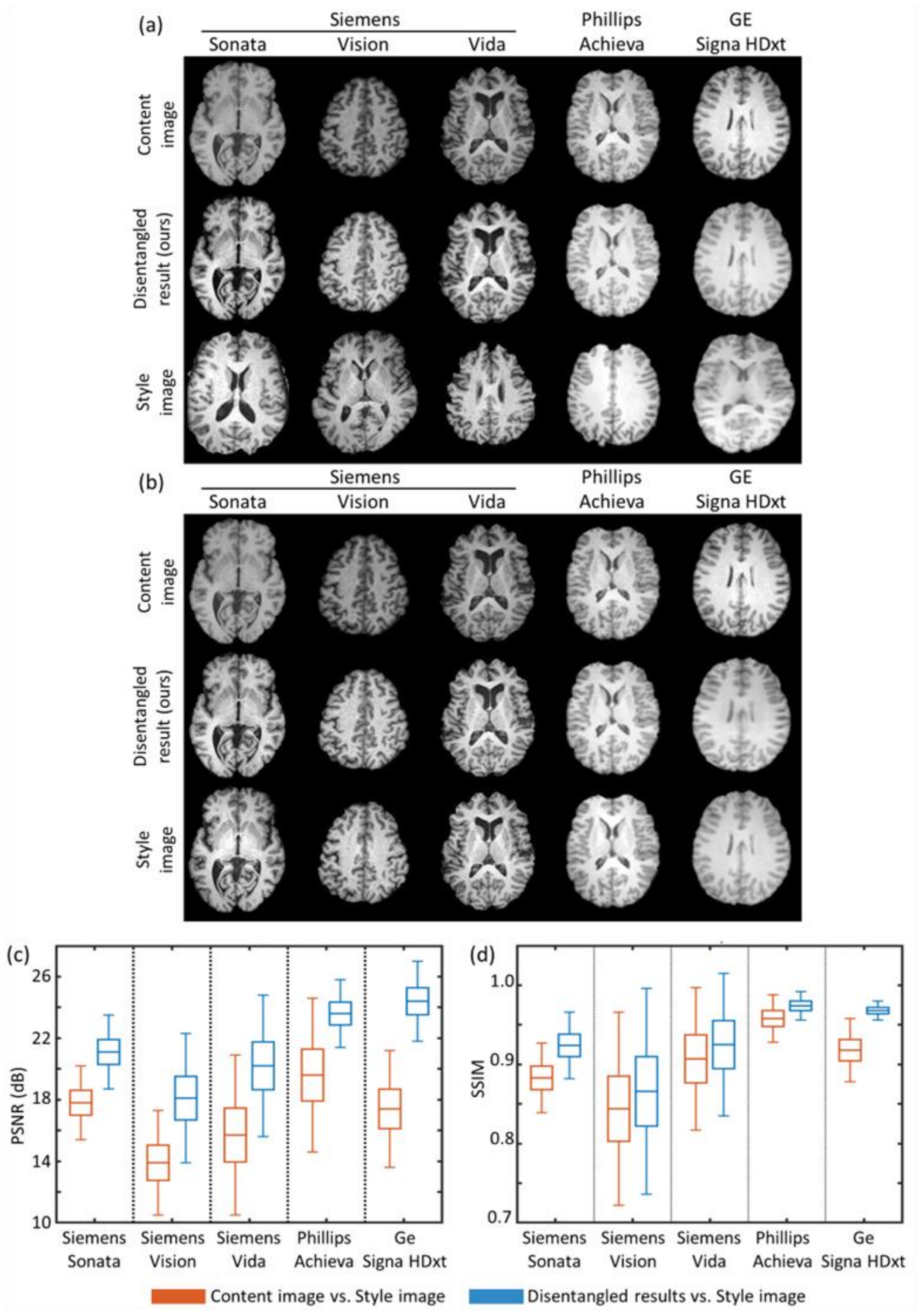

**Figure 2.** Evaluation of the disentanglement performance of the proposed generator. (a) Unpaired setting with both structural and stylistic differences. (b) Paired traveling-subject setting, where structural information is shared but style differs across scanners. Each column shows the content images (top), disentangled results (middle), and style images (bottom). (c)

PSNR and (d) SSIM were measured in the traveling-subjects across scanners, comparing the content images with the corresponding style images (orange boxes) and the disentangled outputs with the style images (blue boxes), respectively. The qualitative and quantitative results demonstrate that the outputs successfully adopt the style characteristics of the style images while preserving the structural information of the content images.

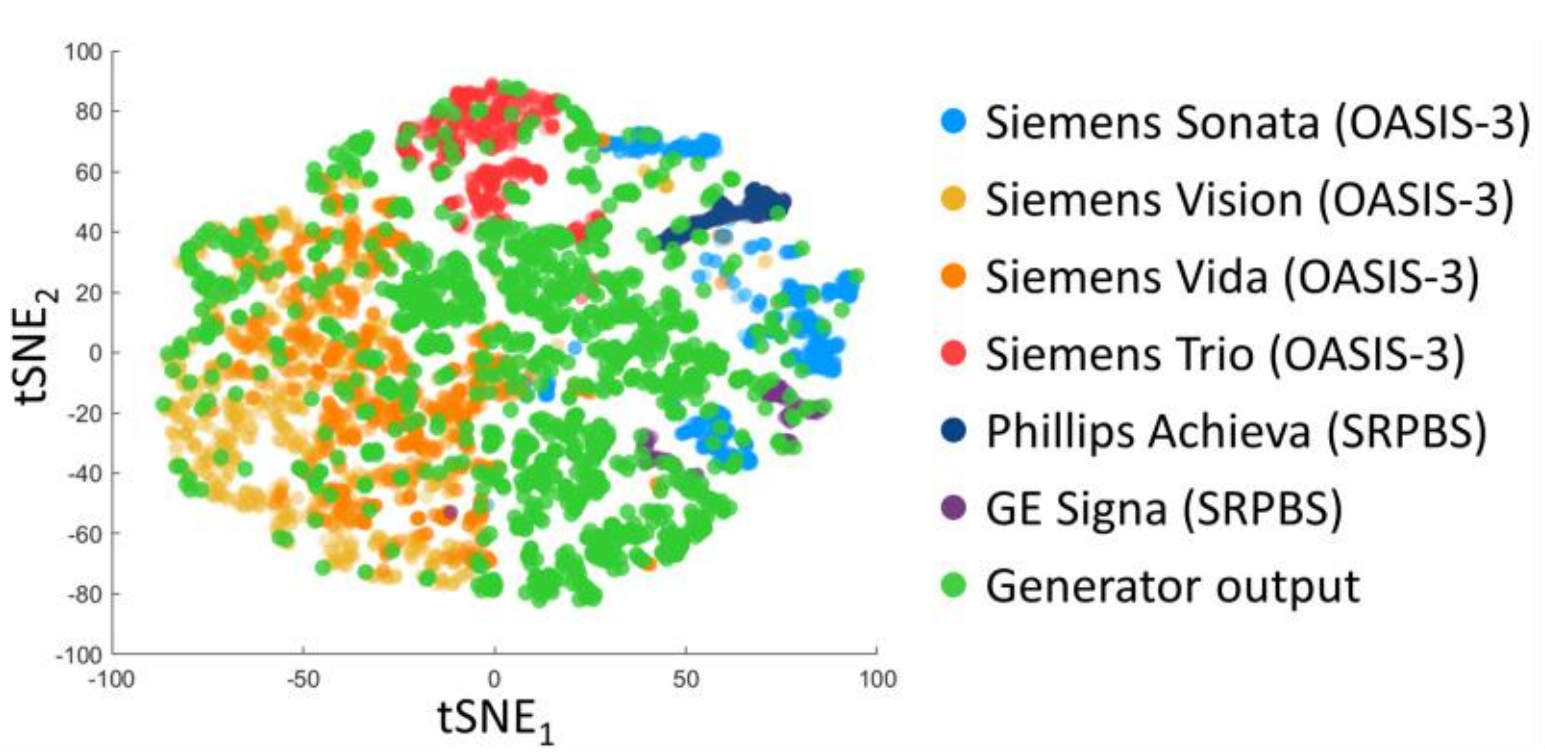

**Figure 3.** t-SNE visualization of MRI images from two open-source datasets (OASIS-3 and SRPBS) and generated images from the proposed generator. Open-source datasets form distinct clusters, while the generated images are more widely dispersed.

## 3. Results

### 3.1. Disentanglement-based Generator Results

Fig. 2(a) presents qualitative results of the style transfer experiment in the unpaired subject setting. The disentangled outputs preserve the anatomical structure of the content images while adopting the appearance of the style images. Fig. 2(b) shows results for the paired subject setting, where the proposed method maintains the structure while adapting the style, demonstrating effective disentanglement. These observations are further supported by quantitative results, where the disentangled outputs achieve higher PSNR and SSIM with respect to the style images than the content images (Fig. 2c).

The experimental results on the inter- and extra-polation of the style vectors show smooth transitions in brightness, contrast, blurriness, and noisiness, while maintaining anatomical structure (Fig. S1 and S2). This provides additional evidence that the generator successfully disentangles style from content.

The t-SNE results of MRI images from the open-source datasets (OASIS-3 and SRPBS) and the generated images are illustrated in Fig. 3. While images from the open-source datasets form domain-specific clusters, the generated images are more broadly distributed in the embedded space, indicating that the generator can produce a wide range of MRI style variations. Qualitative examples of generated images further confirm the diversity of generated image appearances (Fig. S3).

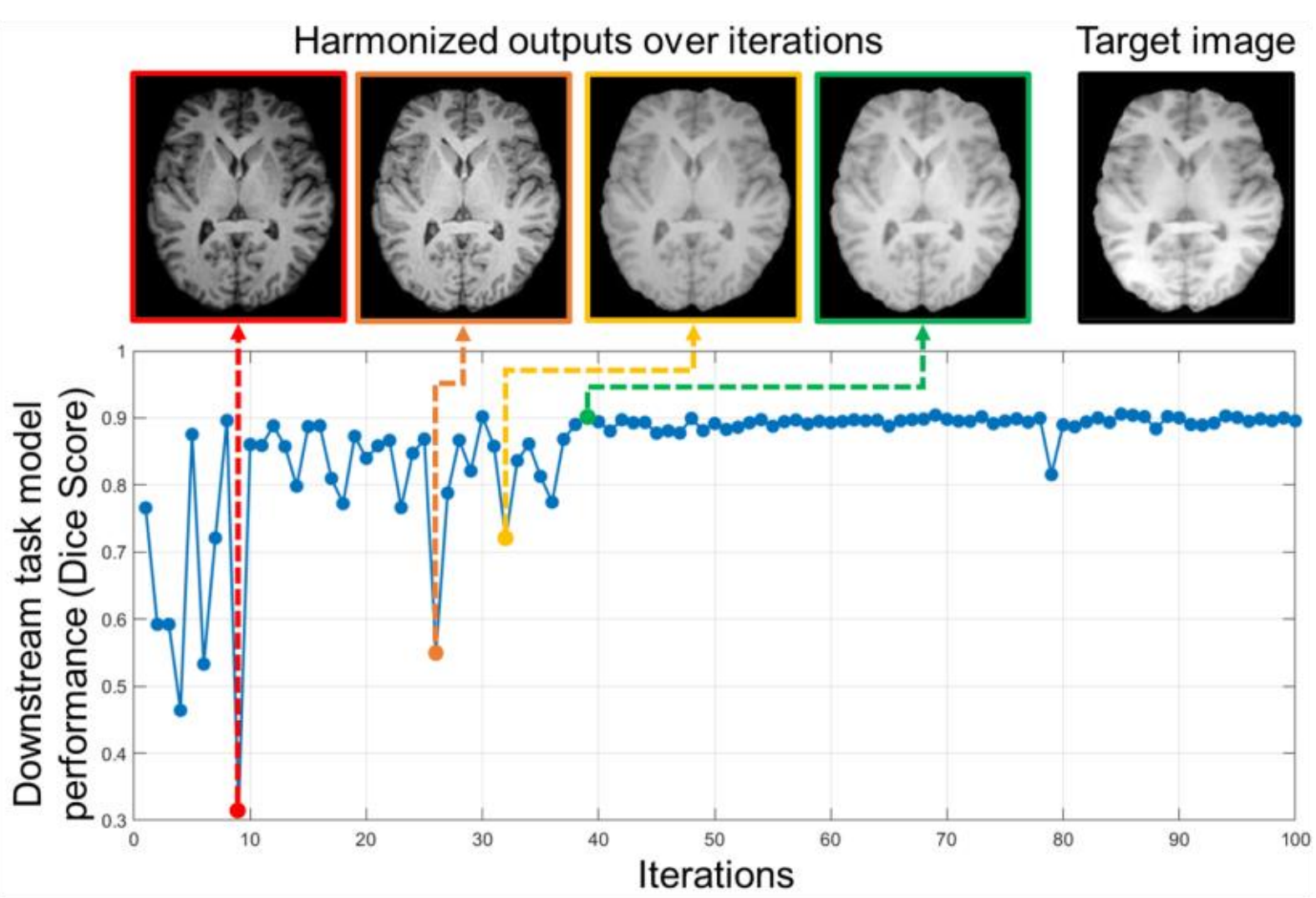

**Figure 4.** Downstream task performance (Dice score) and corresponding harmonized outputs across optimization iterations. As optimization progresses, the performance stabilizes, and the harmonized images gradually resemble the target domain image, demonstrating successful navigation of the generator's style manifold and estimation of the target domain style.

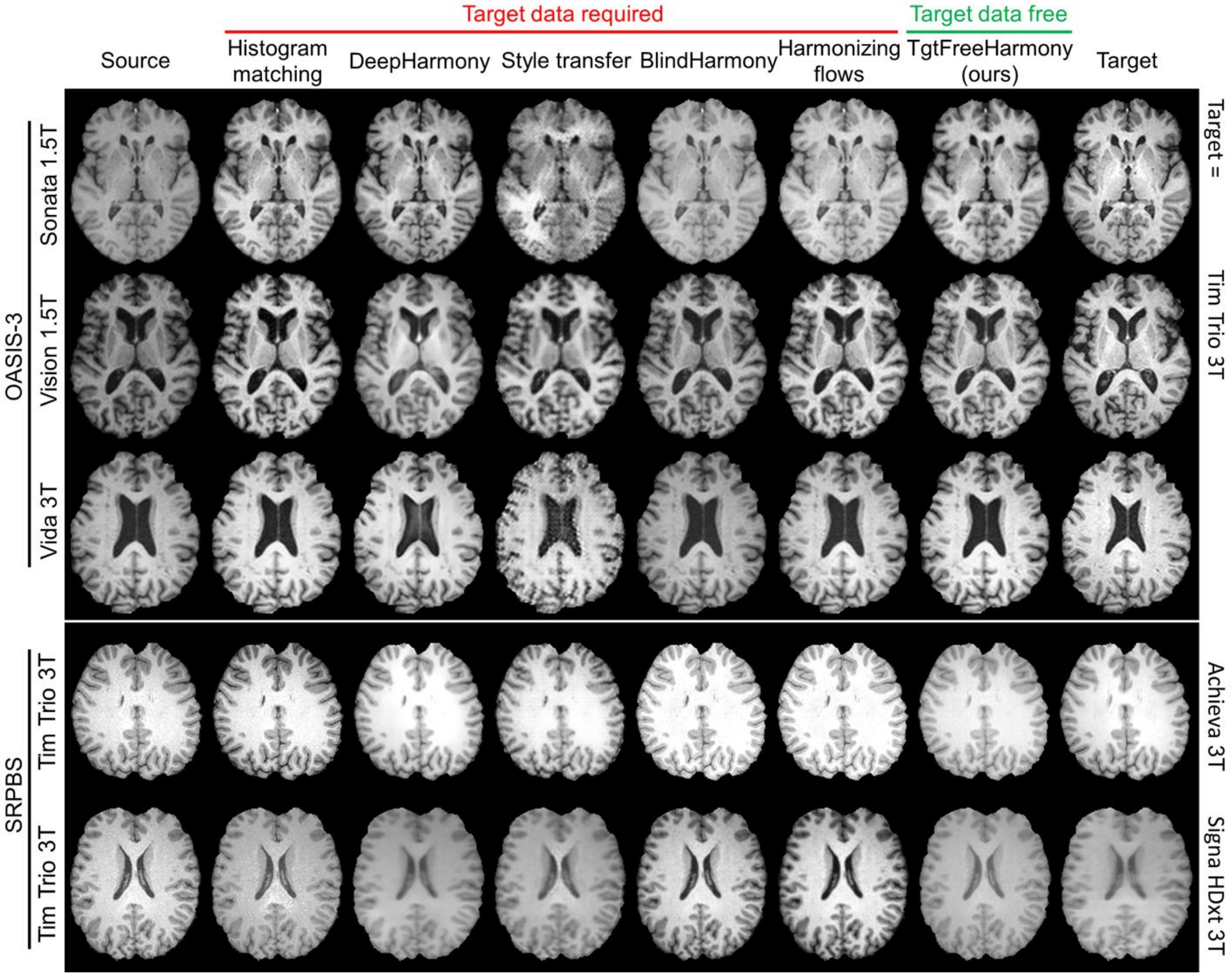

**Figure 5.** Visual comparison of harmonization results across five source-target settings (Siemens Sonata-Tim Trio, Siemens Vision-Tim Trio, Siemens Vida-Tim Trio, Siemens Tim Trio-Philips Achieva, and Siemens Tim Trio-GE Signa HDxt). Methods marked in red require access to target domain data, whereas our method (green) operates without it. For fair comparison, all methods were trained using a single subject. Consequently, methods that rely on learning data distributions (e.g., style transfer, BlindHarmony, and Harmonizing flows) may not reflect their full performance. Histogram matching and DeepHarmony produce visually aligned images with the target, although DeepHarmony exhibits smoothing. Style transfer introduces checkerboard artifacts. In contrast, TgtFreeHarmony achieves close visual alignment with the target images, demonstrating effective harmonization.

**Table 3.** Quantitative comparison of image similarity (PSNR and SSIM) between target and source images before (no harmonization) and after harmonization using different methods (histogram matching, DeepHarmony, style transfer, BlindHarmony, Harmonizing flows, and TgtFreeHarmony) across five source-target settings.

| Dataset | | OASIS-3 | | | | | | SRPBS | | | |
|---|---|---|---|---|---|---|---|---|---|---|---|
| | Target domain requirement | Sonata → Tim Trio | | Vision → Tim Trio | | Magnetom Vida → Tim Trio | | Tim Trio → Achieva | | Tim Trio → Signa HDxt | |
| | | PSNR↑ | SSIM↑ | PSNR↑ | SSIM↑ | PSNR↑ | SSIM↑ | PSNR↑ | SSIM↑ | PSNR↑ | SSIM↑ |
| No harmonization | - | 17.8 ± 1.2 | 0.883 ± 0.022 | 13.9 ± 1.7 | 0.844 ± 0.061 | 15.8 ± 2.4 | 0.887 ± 0.043 | 19.5 ± 2.7 | 0.959 ± 0.015 | 17.4 ± 1.9 | 0.918 ± 0.020 |
| Histogram matching | Required | 20.9 ± 1.2 | **0.924 ± 0.020** | 17.3 ± 1.7 | **0.863 ± 0.063** | **19.5 ± 2.0** | **0.905 ± 0.043** | 22.7 ± 1.0 | 0.960 ± 0.015 | 21.2 ± 1.6 | 0.933 ± 0.015 |
| DeepHarmony | Required | **21.5 ± 2.5** | 0.921 ± 0.025 | **17.8 ± 1.5** | 0.854 ± 0.064 | 19.2 ± 2.0 | 0.901 ± 0.044 | **25.8 ± 1.8** | **0.981 ± 0.006** | **23.4 ± 3.7** | **0.976 ± 0.008** |
| Style transfer | Required | 18.3 ± 1.2 | 0.855 ± 0.032 | 16.5 ± 1.1 | 0.835 ± 0.066 | 16.8 ± 1.2 | 0.861 ± 0.055 | 22.2 ± 1.2 | 0.953 ± 0.012 | 21.2 ± 3.7 | 0.902 ± 0.011 |
| BlindHarmony | Required | 11.2 ± 1.6 | 0.750 ± 0.057 | 11.3 ± 1.8 | 0.765 ± 0.074 | 13.4 ± 1.4 | 0.857 ± 0.063 | 15.7 ± 1.5 | 0.901 ± 0.022 | 15.1 ± 1.9 | 0.896 ± 0.020 |
| Harmonizing flows | Required | 17.6 ± 1.0 | 0.874 ± 0.020 | 16.7 ± 1.7 | 0.854 ± 0.061 | 16.4 ± 2.5 | 0.891 ± 0.046 | 15.7 ± 1.3 | 0.951 ± 0.019 | 16.6 ± 3.4 | 0.909 ± 0.021 |
| TgtFreeHarmony (ours) | Not required | **20.3 ± 1.3** | **0.924 ± 0.020** | **17.5 ± 1.7** | **0.868 ± 0.062** | **17.6 ± 2.1** | **0.896 ± 0.039** | **21.6 ± 1.3** | **0.967 ± 0.009** | **19.4 ± 3.6** | **0.957 ± 0.007** |

### 3.2. Comparison of Harmonization Performance

The downstream segmentation performance (Dice score) across optimization iterations is shown in Fig. 4. The results show that the optimization progressively reduces the variance in downstream task performance while producing harmonized outputs that increasingly resembled the target domain images, characterized by reduced contrast, increased blurriness, and elevated brightness. These results demonstrate that the proposed method successfully navigates the generator's style manifold and estimates the target domain style.

Qualitative results of harmonization are shown in Fig. 5. Histogram matching and DeepHarmony produce visually aligned images relative to the target, though DeepHarmony results appear overly smoothed. Style transfer introduces checkerboard artifacts, whereas source-free harmonization methods, such as BlindHarmony and Harmonizing flows, that rely on domain distribution modeling exhibit failure cases under limited training data conditions. In contrast, TgtFreeHarmony achieves close visual alignment with the target images, demonstrating effective harmonization.

Quantitative results in Table 3 further support these findings. Most harmonization methods improve image similarity metrics to some extent. However, methods that rely on learning data distributions (e.g., style transfer, BlindHarmony, and Harmonizing flows) exhibit inconsistent or failed harmonization in certain domains under extremely limited training data (one subject per source domain; see Discussion). Notably, several deep learning-based methods do not surpass the performance of histogram matching, likely due to the limited training setting where only a single subject was available. The proposed TgtFreeHarmony is the only method that requires no access to target-domain data yet consistently improves image similarity metrics across all evaluated domains.

### 3.3. Comparison of Segmentation Performance After Harmonization

Building upon the harmonization outcomes presented in Section 3.2, we evaluated how each method influenced downstream segmentation performance. Qualitative and quantitative results are summarized in Figure 6 and Table 4, respectively. Most deep learning–based harmonization methods exhibit performance degradation in certain domains, likely due to artifacts or distributional distortions amplified by the limited training data (see Discussion), while histogram matching provides relatively stable improvements in segmentation accuracy across domains. In contrast, TgtFreeHarmony consistently enhances segmentation performance across all evaluated domains.

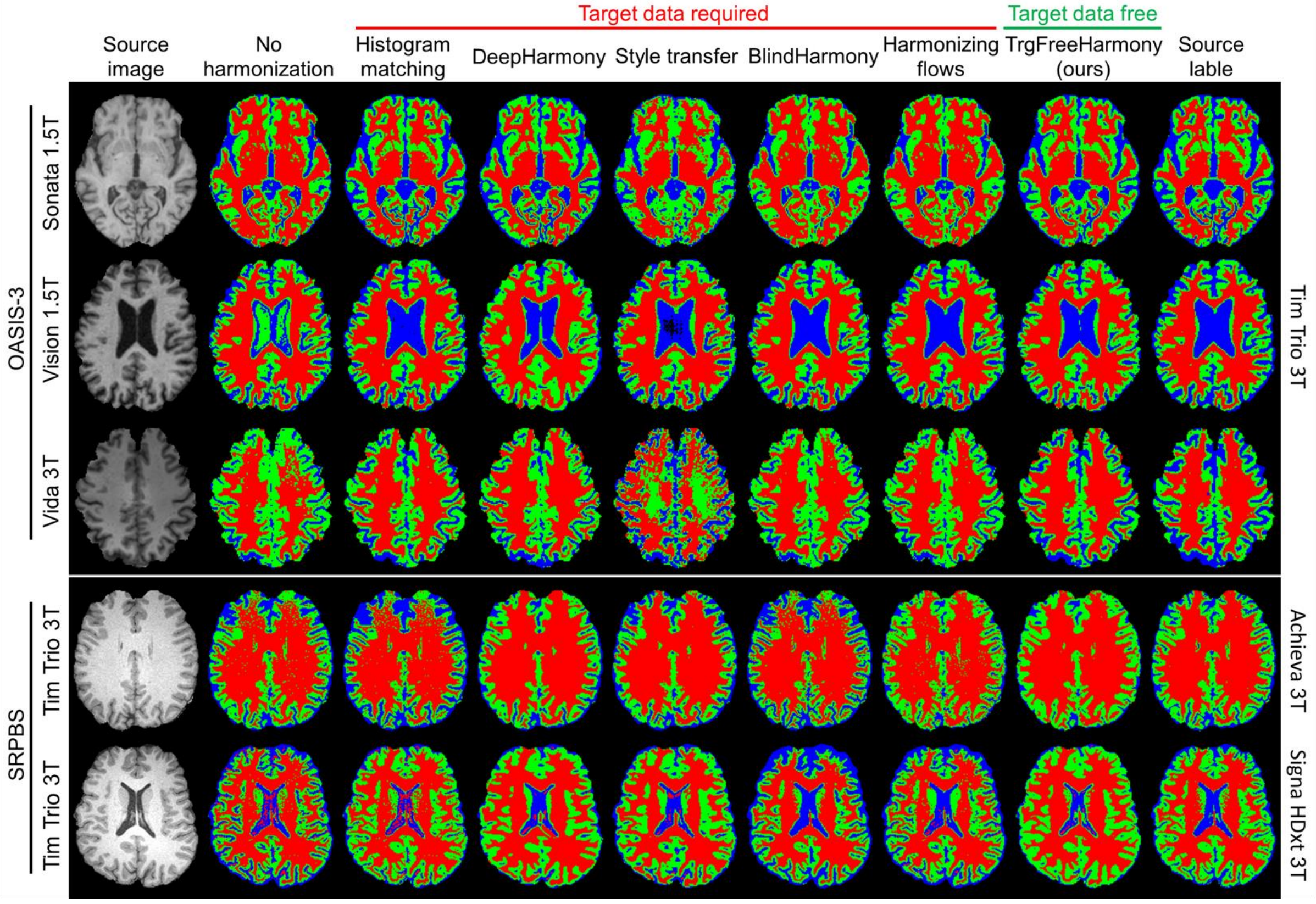

**Figure 6.** Brain tissue segmentation labels and results on source images using different harmonization methods. Without harmonization, segmentation performance degrades due to domain shift. TgtFreeHarmony yields improved brain tissue segmentation across all domains, whereas other deep learning-based harmonization methods, such as Style Transfer, exhibit distortions or artifacts that lead to reduced segmentation performance.

**Table 4.** Quantitative comparison of segmentation metrics (IoU and Dice score) between target and source images before (no harmonization) and after harmonization using different methods (histogram matching, DeepHarmony, style transfer, BlindHarmony, Harmonizing flows, and TgtFreeHarmony) across five source-target settings. For fair comparison, all methods were trained using a single subject; consequently, methods that rely on learning data distributions (e.g., style transfer, BlindHarmony, and Harmonizing flows) may not reflect their full performance as reported in the original studies.

| Dataset | | OASIS-3 | | | | | | SRPBS | | | |
|---|---|---|---|---|---|---|---|---|---|---|---|
| | Target domain requirement | Sonata → Tim Trio | | Vision → Tim Trio | | Magnetom Vida → Tim Trio | | Tim Trio → Achieva | | Tim Trio → Signa HDxt | |
| | | IoU↑ | Dice↑ | IoU↑ | Dice↑ | IoU↑ | Dice↑ | IoU↑ | Dice↑ | IoU↑ | Dice↑ |
| No harmonization | - | 0.711 ± 0.034 | 0.830 ± 0.023 | 0.750 ± 0.067 | 0.852 ± 0.049 | 0.626 ± 0.053 | 0.762 ± 0.045 | 0.563 ± 0.047 | 0.698 ± 0.036 | 0.558 ± 0.042 | 0.692 ± 0.043 |
| Histogram matching | Required | **0.833 ± 0.023** | **0.907 ± 0.014** | **0.818 ± 0.032** | **0.897 ± 0.021** | **0.667 ± 0.045** | **0.788 ± 0.039** | **0.609 ± 0.032** | **0.737 ± 0.026** | **0.658 ± 0.029** | **0.781 ± 0.030** |
| DeepHarmony | Required | 0.709 ± 0.032 | 0.824 ± 0.021 | 0.608 ± 0.061 | 0.751 ± 0.055 | 0.560 ± 0.054 | 0.720 ± 0.050 | 0.559 ± 0.026 | 0.693 ± 0.026 | 0.548 ± 0.024 | 0.691 ± 0.028 |
| Style transfer | Required | 0.665 ± 0.039 | 0.796 ± 0.028 | 0.650 ± 0.040 | 0.785 ± 0.031 | 0.490 ± 0.043 | 0.669 ± 0.030 | 0.520 ± 0.034 | 0.682 ± 0.032 | 0.635 ± 0.030 | 0.767 ± 0.029 |
| BlindHarmony | Required | 0.572 ± 0.047 | 0.711 ± 0.021 | 0.624 ± 0.044 | 0.720 ± 0.028 | 0.471 ± 0.049 | 0.614 ± 0.038 | 0.518 ± 0.031 | 0.677 ± 0.039 | 0.520 ± 0.035 | 0.674 ± 0.022 |
| Harmonizing flows | Required | 0.682 ± 0.048 | 0.809 ± 0.034 | 0.810 ± 0.049 | 0.892 ± 0.034 | 0.638 ± 0.047 | 0.770 ± 0.040 | 0.540 ± 0.025 | 0.685 ± 0.025 | 0.555 ± 0.047 | 0.691 ± 0.046 |
| TgtFreeHarmony (ours) | Not required | **0.839 ± 0.024** | **0.911 ± 0.015** | **0.822 ± 0.045** | **0.899 ± 0.030** | **0.682 ± 0.040** | **0.806 ± 0.030** | **0.684 ± 0.031** | **0.806 ± 0.025** | **0.704 ± 0.065** | **0.820 ± 0.060** |

## 4. Discussion

In this study, we introduced TgtFreeHarmony, a novel target-free harmonization framework, designed for scenarios in which target domain data are inaccessible. By leveraging a disentanglement-based generator, the method separated the domain-variant style of the image from the domain-invariant anatomical content, enabling harmonization to be performed by modifying only the style component (Fig. 2). Notably, the method demonstrated effective style transfer even on unseen domains (Siemens Tim Trio, Philips Achieva, and GE Signa HDxt), indicating promising generalization beyond the training domains. The interpolation and extrapolation experiments further validated successful disentanglement, showing continuous transitions in brightness, contrast, blurriness, and noisiness while preserving anatomical structures (see Section S2).

The diversity observed in both the feature space via t-SNE (Fig. 3) and the image space through generated samples (Fig. S3) demonstrated that the generator's style manifold effectively captures a broad range of MRI appearance variations.

Despite using only a single subject for the optimization, TgtFreeHarmony successfully estimated target styles within the generator's latent space guided by downstream task performance. In contrast, most deep learning–based harmonization methods struggled with this setting of extremely limited data (Figs. 5 and 6, and Tables 3 and 4). For instance, unsupervised style transfer methods, often implemented using generative adversarial networks (Zhu et al., 2017), produced checkerboard artifacts under limited training data, which degraded the performance of the downstream segmentation task (Karras et al., 2020). Source-free methods such as BlindHarmony and Harmonizing flows showed inconsistent performance, as they require substantial data to model source-domain distributions, and with an insufficient number of training subjects, they failed to adequately capture domain variability (Beizaee et al., 2025; Jeong et al., 2023). DeepHarmony, which leverages paired data for voxel-wise mapping, is highly sensitive to registration quality. Since perfect registration is difficult to achieve, this sensitivity introduces a risk of unintended structural alterations during harmonization, which can degrade segmentation performance. The observed improvement in SSIM is likely attributable to enhanced luminance and contrast, as SSIM reflects not only structural similarity but also these appearance-related factors. In contrast, histogram matching, a non-deep learning baseline, achieved strong performance across all domains, highlighting the challenges faced by deep learning-based harmonization in severely data-limited settings. Notably, TgtFreeHarmony estimates the target domain using sampled style vectors within the generator's style manifold, making it less sensitive to the number of training subjects, as the optimization does not rely on direct data-driven mapping. Since the proposed method requires labeled source data, one option is to bypass harmonization and fine-tune the downstream task model directly on the source domain. However, experiments have shown that training segmentation models directly on such limited data leads to severe overfitting and degraded performance (see Section S3). These results underscore the advantage of TgtFreeHarmony in settings where labeled data is scarce, as is common in clinical environments.

One potential limitation of harmonization methods that operate on 2D slices is the introduction of slice discontinuities when the methods are applied to whole-brain volumes. Although the proposed method also processes images in a 2D slice-wise manner, this issue is mitigated by applying a single estimated style vector consistently across all slices and subjects within a source domain. Consequently, the harmonized volumes exhibit a coherent appearance without noticeable slice discontinuity (Fig. S4).

Despite these strengths, TgtFreeHarmony has several limitations. First, the experimental evaluation is restricted to a limited set of domains from the available open-source datasets and may not reflect the full diversity of real-world MRI protocols. The current framework also lacks an explicit mechanism for quantifying the coverage of the learned style manifold. Future work should examine robustness across a broader range of imaging conditions, such as acquisition sequences (e.g., T2-weighted). Furthermore, although TgtFreeHarmony requires only a small number of labels, it still incurs an annotation cost. The performance of the method also depends on the quality of the downstream task model, as the optimization process is guided by the downstream model's performance. Consequently, any bias or suboptimal generalization of the downstream model may directly affect the estimated target style and the resulting harmonization. Another limitation lies in the assumption of a single representative style per source domain, achieved by estimating a single optimal style vector.

This assumption may be insufficient to capture intra-domain variability, particularly in scenarios where multiple acquisition conditions exist within a domain. In addition, the method may not generalize effectively to all downstream tasks. For tasks with discrete performance metrics, such as classification, the metrics may not provide sufficiently informative guidance for harmonization. Lastly, most experiments involved healthy subjects. It remains unclear whether the generator preserves clinically relevant features when applied to pathological data, such as lesions. Validation on more diverse and pathological datasets will be essential to establish the clinical reliability and applicability of TgtFreeHarmony.

## 5. Conclusion

We presented TgtFreeHarmony, a harmonization framework that operates without access to the target domain. The method leverages disentangled representation learning to construct an MRI style manifold that captures domain-specific variations while preserving anatomical content. Guided by downstream task performance, TgtFreeHarmony efficiently estimates the target domain style within the generator's style manifold and performs harmonization accordingly. This approach broadens the applicability of harmonization in real-world clinical environments under strict data-sharing constraints.

## Acknowledgments

This research was supported by Korean Government Grants: RS-2024-00439677, IITP-2026-RS-2023-00256081, RS-2024-00435727, and Institute of New Media and Communications and Institute of Engineering Research, and Electric Power Research Institute of Seoul National University, and Samsung Electronics Co., Ltd (IO260313-15905-01).

## Conflict of interest statement

The authors have no conflicts of interest to disclose.

## Data availability statement

The data used in this study are publicly available open-source datasets. The code has been openly available at https://github.com/SNU-LIST/TgtFreeHarmony.git.

## References

Beizaee, F., Lodygensky, G.A., Adamson, C.L., Thompson, D.K., Cheong, J.L., Spittle, A.J., Anderson, P.J., Desrosiers, C., Dolz, J., 2025. Harmonizing flows: Leveraging normalizing flows for unsupervised and source-free MRI harmonization. Medical Image Analysis 101, 103483.

Ben-David, S., Blitzer, J., Crammer, K., Pereira, F., 2006. Analysis of representations for domain adaptation. Advances in neural information processing systems 19.

Billot, B., Magdamo, C., Cheng, Y., Arnold, S.E., Das, S., Iglesias, J.E., 2023. Robust machine learning segmentation for large-scale analysis of heterogeneous clinical brain MRI datasets. Proceedings of the National Academy of Sciences 120, e2216399120.

Brochu, E., Cora, V.M., De Freitas, N., 2010. A tutorial on Bayesian optimization of expensive cost functions, with application to active user modeling and hierarchical reinforcement learning. arXiv preprint arXiv:1012.2599.

Cackowski, S., Barbier, E.L., Dojat, M., Christen, T., 2023. ImUnity: A generalizable VAE-GAN solution for multicenter MR image harmonization. Medical Image Analysis 88, 102799.

Cai, L.Y., Yang, Q., Kanakaraj, P., Nath, V., Newton, A.T., Edmonson, H.A., Luci, J., Conrad, B.N., Price, G.R., Hansen, C.B., 2021. MASiVar: Multisite, multiscanner, and multisubject acquisitions for studying variability in diffusion weighted MRI. Magnetic resonance in medicine 86, 3304–3320.

Chen, X., Duan, Y., Houthooft, R., Schulman, J., Sutskever, I., Abbeel, P., 2016. Infogan: Interpretable representation learning by information maximizing generative adversarial nets. Advances in neural information processing systems 29.

Cui, Z., Li, C., Du, Z., Chen, N., Wei, G., Chen, R., Yang, L., Shen, D., Wang, W., 2021. Structure-driven unsupervised domain adaptation for cross-modality cardiac segmentation. IEEE Transactions on Medical Imaging 40, 3604–3616.

Dewey, B.E., Zhao, C., Reinhold, J.C., Carass, A., Fitzgerald, K.C., Sotirchos, E.S., Saidha, S., Oh, J., Pham, D.L., Calabresi, P.A., 2019. DeepHarmony: A deep learning approach to contrast harmonization across scanner changes. Magnetic resonance imaging 64, 160–170.

Dewey, B.E., Zuo, L., Carass, A., He, Y., Liu, Y., Mowry, E.M., Newsome, S., Oh, J., Calabresi, P.A., Prince, J.L., 2020. A disentangled latent space for cross-site MRI harmonization, International conference on medical image computing and computer-assisted intervention. Springer, pp. 720–729.

Frazier, P.I., 2018. A tutorial on Bayesian optimization. arXiv preprint arXiv:1807.02811.

Fredrikson, M., Jha, S., Ristenpart, T., 2015. Model inversion attacks that exploit confidence information and basic countermeasures, Proceedings of the 22nd ACM SIGSAC conference on computer and communications security, pp. 1322–1333.

Gardner, J., Pleiss, G., Weinberger, K.Q., Bindel, D., Wilson, A.G., 2018. Gpytorch: Blackbox matrix-matrix gaussian process inference with gpu acceleration. Advances in neural information processing systems 31.
Guan, H., Liu, Y., Yang, E., Yap, P.-T., Shen, D., Liu, M., 2021. Multi-site MRI harmonization via attention-guided deep domain adaptation for brain disorder identification. Medical image analysis 71, 102076.
He, K., Zhang, X., Ren, S., Sun, J., 2016. Deep residual learning for image recognition, Proceedings of the IEEE conference on computer vision and pattern recognition, pp. 770–778.
Huang, X., Belongie, S., 2017. Arbitrary style transfer in real-time with adaptive instance normalization, Proceedings of the IEEE international conference on computer vision, pp. 1501–1510.
Huang, X., Liu, M.-Y., Belongie, S., Kautz, J., 2018. Multimodal unsupervised image-to-image translation, Proceedings of the European conference on computer vision (ECCV), pp. 172–189.
Jenkinson, M., Bannister, P., Brady, M., Smith, S., 2002. Improved optimization for the robust and accurate linear registration and motion correction of brain images. Neuroimage 17, 825–841.
Jeong, H., Byun, H., Kang, D.U., Lee, J., 2023. BlindHarmony:" blind" harmonization for MR images via flow model, Proceedings of the IEEE/CVF International Conference on Computer Vision, pp. 21129–21139.
Karras, T., Aittala, M., Hellsten, J., Laine, S., Lehtinen, J., Aila, T., 2020. Training generative adversarial networks with limited data. Advances in neural information processing systems 33, 12104–12114.
Kingma, D.P., 2014. Adam: A method for stochastic optimization. arXiv preprint arXiv:1412.6980.
Kornblith, S., Shlens, J., Le, Q.V., 2019. Do better imagenet models transfer better?, Proceedings of the IEEE/CVF conference on computer vision and pattern recognition, pp. 2661–2671.
Kushol, R., Wilman, A.H., Kalra, S., Yang, Y.-H., 2023. DSMRI: domain shift analyzer for multi-center MRI datasets. Diagnostics 13, 2947.
LaMontagne, P.J., Benzinger, T.L., Morris, J.C., Keefe, S., Hornbeck, R., Xiong, C., Grant, E., Hassenstab, J., Moulder, K., Vlassenko, A.G., 2019. OASIS-3: longitudinal neuroimaging, clinical, and cognitive dataset for normal aging and Alzheimer disease. medrxiv, 2019.2012. 2013.19014902.
Li, C., Wong, C., Zhang, S., Usuyama, N., Liu, H., Yang, J., Naumann, T., Poon, H., Gao, J., 2023. Llava-med: Training a large language-and-vision assistant for biomedicine in one day. Advances in Neural Information Processing Systems 36, 28541–28564.
Liu, M., Maiti, P., Thomopoulos, S., Zhu, A., Chai, Y., Kim, H., Jahanshad, N., 2021. Style transfer using generative adversarial networks for multi-site MRI harmonization, International conference on medical image computing and computer-assisted intervention. Springer, pp. 313–322.
Liu, S., Yap, P.-T., 2024. Learning multi-site harmonization of magnetic resonance images without traveling human phantoms. Communications Engineering 3, 6.
Long, M., Cao, Y., Wang, J., Jordan, M., 2015. Learning transferable features with deep adaptation networks, International conference on machine learning. PMLR, pp. 97–105.
Ma, J., He, Y., Li, F., Han, L., You, C., Wang, B., 2024. Segment anything in medical images. Nature Communications 15, 654.
Modanwal, G., Vellal, A., Buda, M., Mazurowski, M.A., 2020. MRI image harmonization using cycle-consistent generative adversarial network, Medical Imaging 2020: Computer-Aided Diagnosis. SPIE, pp. 259–264.
Moor, M., Huang, Q., Wu, S., Yasunaga, M., Dalmia, Y., Leskovec, J., Zakka, C., Reis, E.P., Rajpurkar, P., 2023. Med-flamingo: a multimodal medical few-shot learner, Machine Learning for Health (ML4H). PMLR, pp. 353–367.
Nyúl, L.G., Udupa, J.K., Zhang, X., 2000. New variants of a method of MRI scale standardization. IEEE transactions on medical imaging 19, 143–150.
Roca, V., Kuchcinski, G., Pruvo, J.-P., Manouvriez, D., Lopes, R., 2025. IGUANe: A 3D generalizable CycleGAN for multicenter harmonization of brain MR images. Medical Image Analysis 99, 103388.
Ronneberger, O., Fischer, P., Brox, T., 2015. U-net: Convolutional networks for biomedical image segmentation, Medical Image Computing and Computer-Assisted Intervention–MICCAI 2015: 18th International Conference, Munich, Germany, October 5-9, 2015, Proceedings, Part III 18. Springer, pp. 234–241.

Scholz, D., Erdur, A.C., Holland, R., Ehm, V., Peeken, J.C., Wiestler, B., Rueckert, D., 2025. Contrastive Anatomy-Contrast Disentanglement: A Domain-General MRI Harmonization Method, International Conference on Medical Image Computing and Computer-Assisted Intervention. Springer, pp. 100–110.

Shinohara, R.T., Sweeney, E.M., Goldsmith, J., Shiee, N., Mateen, F.J., Calabresi, P.A., Jarso, S., Pham, D.L., Reich, D.S., Crainiceanu, C.M., 2014. Statistical normalization techniques for magnetic resonance imaging. NeuroImage: Clinical 6, 9–19.

Srinivas, N., Krause, A., Kakade, S.M., Seeger, M., 2009. Gaussian process optimization in the bandit setting: No regret and experimental design. arXiv preprint arXiv:0912.3995.

Tajbakhsh, N., Shin, J.Y., Gurudu, S.R., Hurst, R.T., Kendall, C.B., Gotway, M.B., Liang, J., 2016. Convolutional neural networks for medical image analysis: Full training or fine tuning? IEEE transactions on medical imaging 35, 1299–1312.

Tanaka, S.C., Yamashita, A., Yahata, N., Itahashi, T., Lisi, G., Yamada, T., Ichikawa, N., Takamura, M., Yoshihara, Y., Kunimatsu, A., 2021. A multi-site, multi-disorder resting-state magnetic resonance image database. Scientific data 8, 227.

Ulyanov, D., Vedaldi, A., Lempitsky, V., 2016. Instance normalization: The missing ingredient for fast stylization. arXiv preprint arXiv:1607.08022.

Wang, M., Deng, W., 2018. Deep visual domain adaptation: A survey. Neurocomputing 312, 135–153.

Williams, C.K., Rasmussen, C.E., 2006. Gaussian processes for machine learning. MIT press Cambridge, MA.

Zhang, Y., Jia, R., Pei, H., Wang, W., Li, B., Song, D., 2020. The secret revealer: Generative model-inversion attacks against deep neural networks, Proceedings of the IEEE/CVF conference on computer vision and pattern recognition, pp. 253–261.

Zhu, J.-Y., Park, T., Isola, P., Efros, A.A., 2017. Unpaired image-to-image translation using cycle-consistent adversarial networks, Proceedings of the IEEE international conference on computer vision, pp. 2223–2232.

Zuo, L., Dewey, B.E., Carass, A., Liu, Y., He, Y., Calabresi, P.A., Prince, J.L., 2021a. Information-based disentangled representation learning for unsupervised MR harmonization, International Conference on Information Processing in Medical Imaging. Springer, pp. 346–359.

Zuo, L., Dewey, B.E., Liu, Y., He, Y., Newsome, S.D., Mowry, E.M., Resnick, S.M., Prince, J.L., Carass, A., 2021b. Unsupervised MR harmonization by learning disentangled representations using information bottleneck theory. NeuroImage 243, 118569.

# Supplementary information

## 1. Downstream segmentation task network architecture

The downstream segmentation task network was implemented as a 2D U-net (Ronneberger et al., 2015), consisting of an encoder–decoder architecture with symmetric skip connections. The encoder comprises four downsampling stages, each with two convolutional layers followed by a pooling layer. The decoder mirrors this structure with four upsampling layers, each consisting of an upsampling operation followed by two convolutional layers, where features are concatenated with corresponding encoder features via skip connections.

## 2. Additional Analysis of Disentanglement-based Generator

To further investigate the behavior of the learned latent space of the generator, we conducted interpolation and extrapolation experiments between style vectors extracted from different MRI images. Given two style vectors, a weighted combination was formed using predefined weights, where one vector was scaled by values $w$ = [-0.5, -0.25, 0, 0.33, 0.66, 1, 1.25, 1.5] and the other by the complementary weights $1 - w$. By varying these weights, both interpolation (within the range [0, 1]) and extrapolation (outside this range) were performed. These experiments were performed not only within the seen domains used for generator training (Siemens Vision and Vida) but also between seen and unseen domains (Siemens Sonata and Tim Trio). Across all cases, the synthesized images exhibited smooth and continuous transitions in appearance while maintaining anatomical structure, demonstrating effective disentanglement of content and style (Chen et al., 2016) (Fig. S1).

We additionally examined the diversity of appearance variations represented in the latent space using synthetically generated image pairs. For a given MRI image, we created synthetic style-shifted counterparts by applying controlled modifications along four appearance dimensions, which are intensity, contrast, blurriness, and noisiness. All counterparts were created from input images normalized to the range [0, 1]. Intensity was adjusted by scaling the image by 1.3 and adding an offset of 0.25. Contrast was altered through gamma correction with a gamma value of 0.7. Blurriness was introduced using the addition of Gaussian blur with a standard deviation of 0.5, and noisiness was simulated by adding Gaussian with a standard noise deviation of 0.1. Interpolation and extrapolation between the original and synthetically altered style vectors revealed continuous and directionally consistent changes in each appearance component, further confirming that the latent space captures a broad and semantically meaningful spectrum of MRI style variations (Fig. S2).

The proposed generator enables the synthesis of MRI images with diverse styles, where each style can be viewed as representing a different domain. Fig. S3 illustrates various generated images by combining randomly sampled style vectors with a fixed content vector.

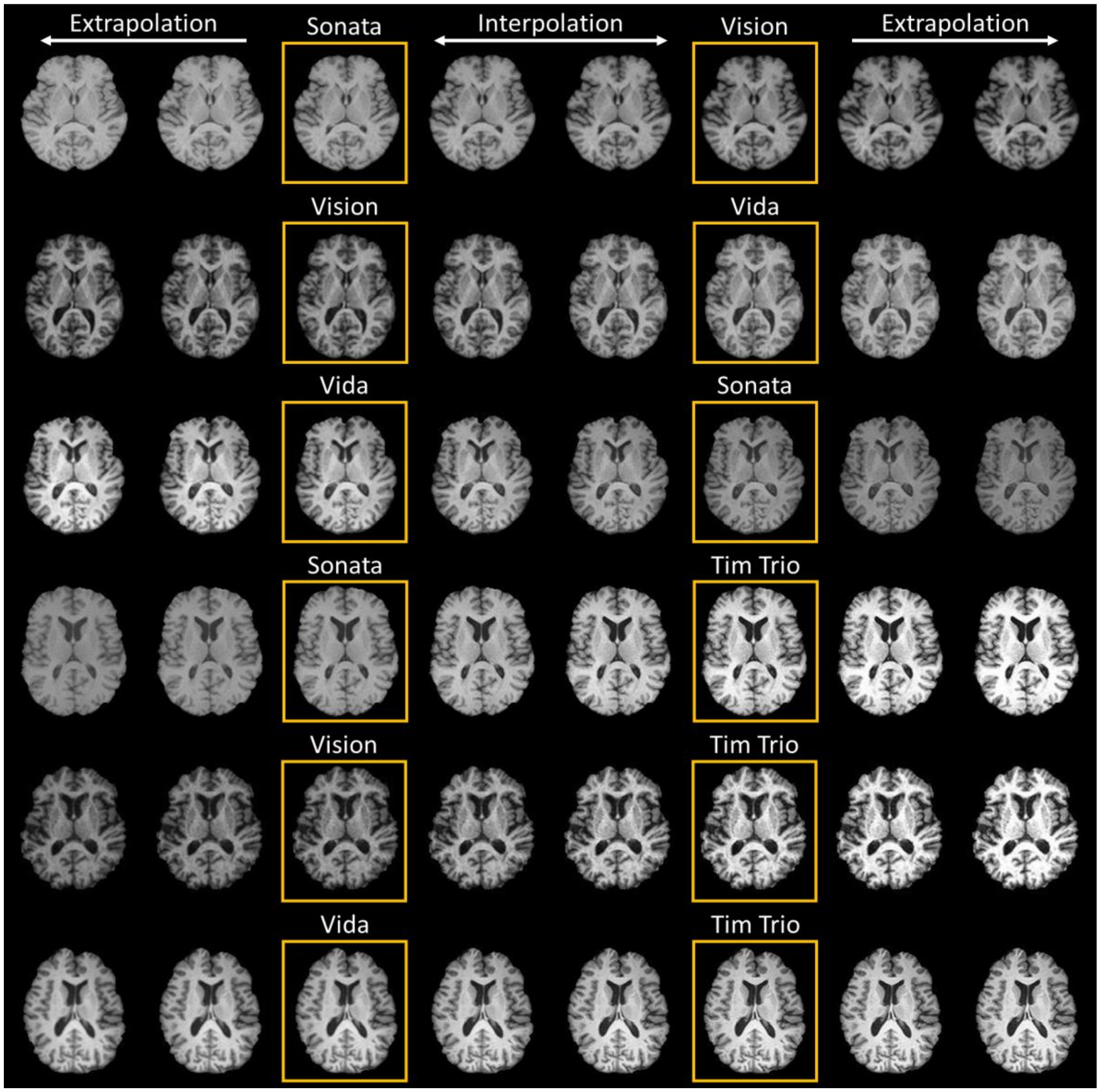

**Figure S1.** Interpolation and extrapolation in the style latent space. Style vectors extracted from two images (highlighted in yellow boxes) were interpolated and extrapolated to synthesize outputs. The top three rows present results using style vectors from domains included in generator training (Siemens Vision and Vida), whereas the bottom three rows illustrate transitions between seen and an unseen domain (Siemens Sonata and Tim Trio). The generated images exhibit smooth, continuous change in appearance while preserving anatomical structure, demonstrating effective disentanglement of the MRI image's content and style.

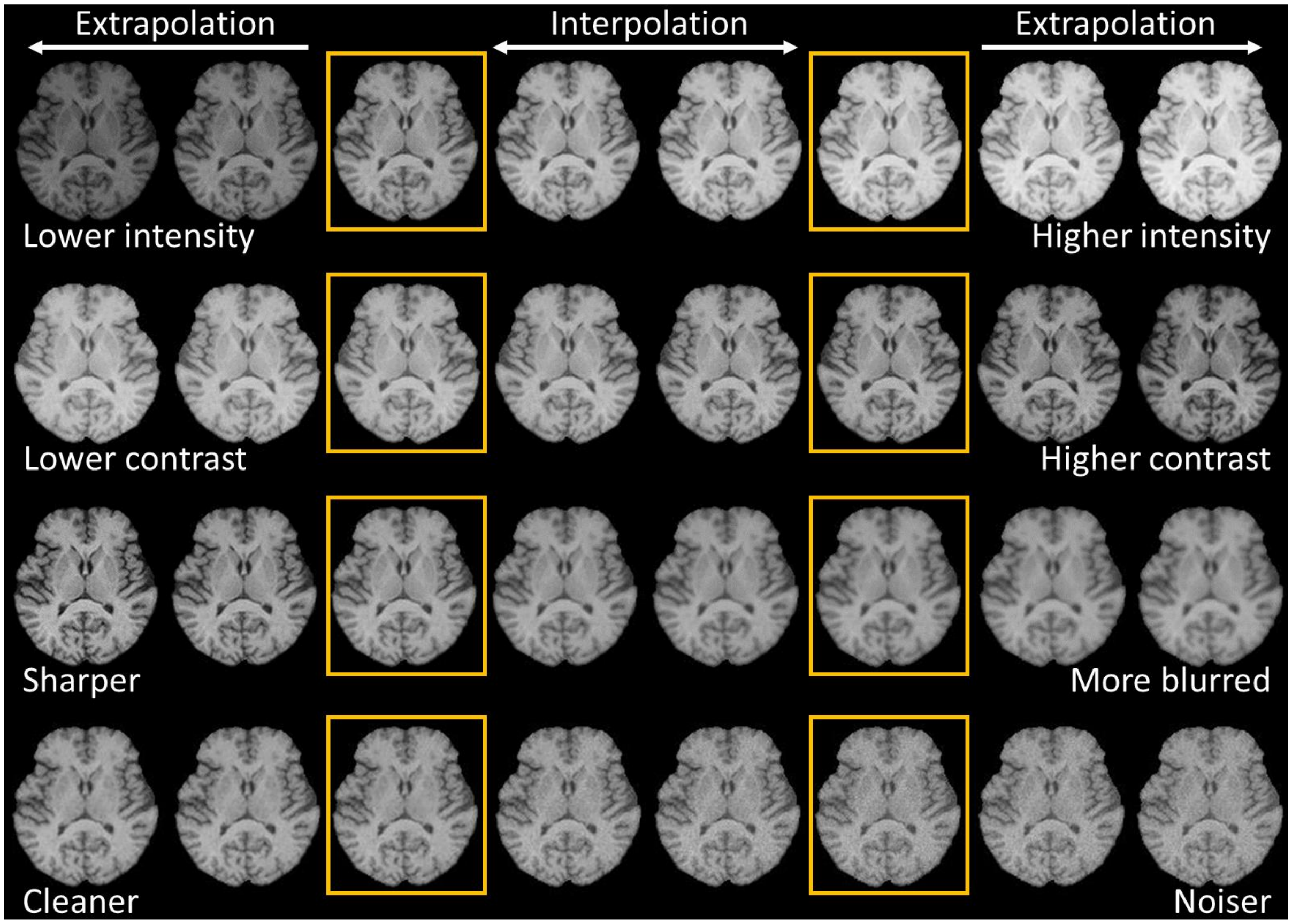

**Figure S2.** MRI images generated with controlled variations in brightness, contrast, blurring, and noisiness. For each attribute, a perturbed version of the original image was created by modifying only the corresponding property. Style vectors extracted from the original and perturbed images (marked with yellow boxes) were then interpolated and extrapolated to synthesize outputs. The results exhibit smooth, attribute-specific transitions while maintaining anatomical consistency, demonstrating that the proposed method captures a broad and coherent spectrum of MRI appearance variations.

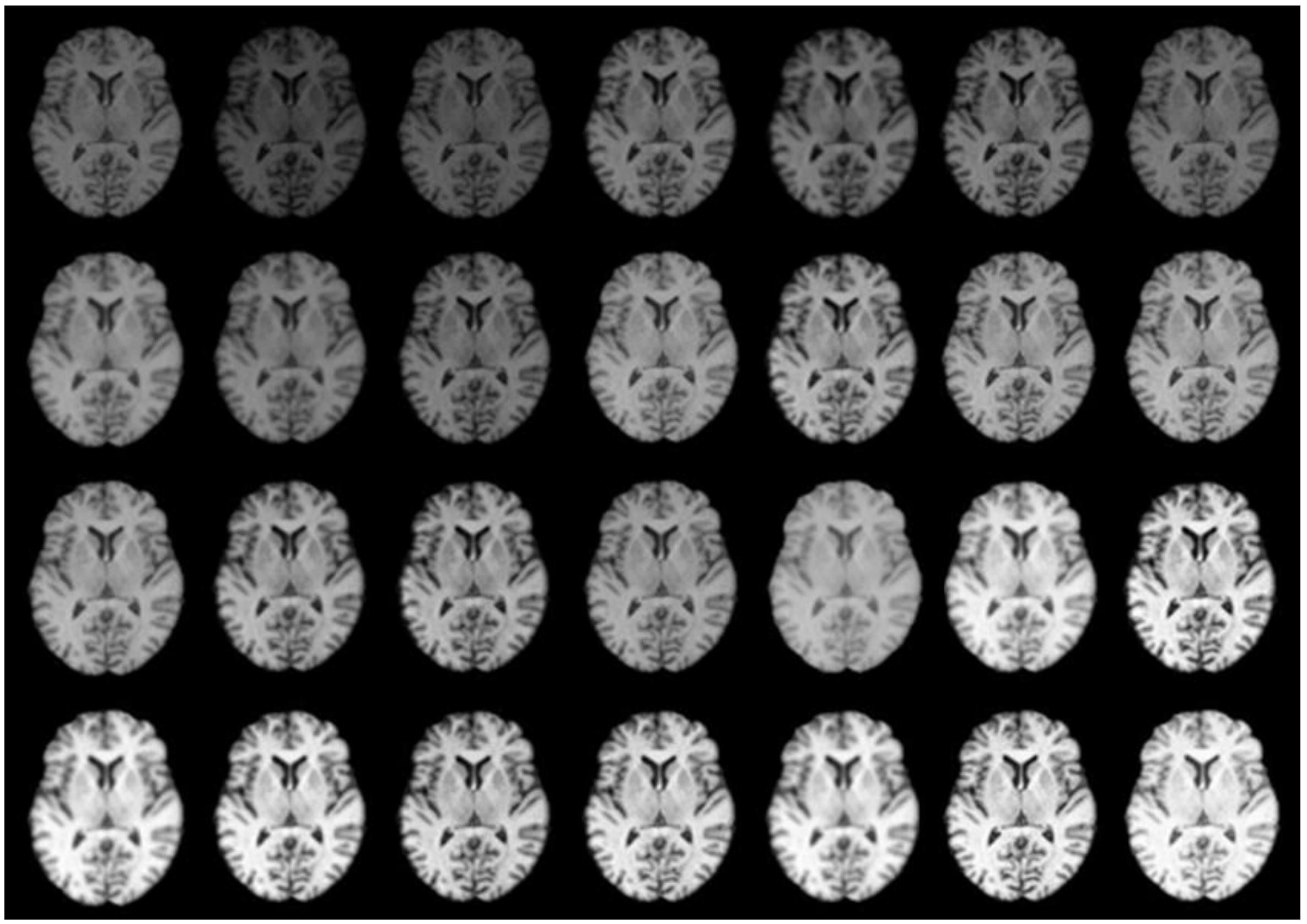

**Figure S3.** MRI images generated by the disentanglement-based generator with the fixed content feature and randomly sampled style vectors. The generated images show preserved anatomical structure while varying image appearance.

## 3. Additional Comparison with Supervised Learning on Limited Labels

Although our method does not require access to target domain data, it assumes the availability of a small number of labeled source domain data. In such a scenario, one might question whether directly supervised training or adapting fine-tuning on the labeled target data could outperform our harmonization approach. To investigate this, we conducted two additional comparisons on OASIS-3, where the dataset size is sufficiently large to support stable supervised training. First, we trained U-Net (Ronneberger et al., 2015) for brain tissue segmentation in a fully supervised manner using varying numbers of labeled subjects (1, 5, 10, 20, and 40). As shown in Table S1, models trained with only 1 labeled subject, matching our harmonization setting, consistently underperformed relative to TgtFreeHarmony. Competitive performance emerges only after substantially increasing the labeled training size to over 20 subjects. Second, we evaluated a fine-tuning baseline in which the segmentation model pretrained on the rich target domain was further trained using the available labeled source domain subjects (1, 5, 10, and 20). As summarized in Table S2, this fine-tuning strategy failed under our setting (1 subject only) and required considerably more labeled data, with some domains needing at least five subjects and others requiring more than ten to approach the performance achieved by our harmonization method. These results highlight the risk of overfitting with small datasets and the practicality of harmonization in settings where labeled data is scarce. Our method is especially beneficial in clinical environments, where obtaining labels typically requires expert knowledge and high costs.

**Table S1.** Segmentation performance comparison between our method (TgtFreeHarmony) and fully supervised models trained with varying numbers of labeled subjects (1, 5, 10, 20, and 40) from each source domain. Scores that surpass TgtFreeHarmony are underlined.

| Dataset | OASIS-3 | | | | | |
|---|---|---|---|---|---|---|
| | Sonata → Tim Trio | | Vision → Tim Trio | | Magnetom Vida → Tim Trio | |
| | IoU↑ | Dice↑ | IoU↑ | Dice↑ | IoU↑ | Dice↑ |
| Supervised model (1 subjects) | 0.494 ± 0.112 | 0.631 ± 0.106 | 0.744 ± 0.094 | 0.835 ± 0.083 | 0.546 ± 0.045 | 0.684 ± 0.046 |
| Supervised model (5 subjects) | 0.669 ± 0.019 | 0.790 ± 0.018 | 0.758 ± 0.084 | 0.841 ± 0.075 | 0.689 ± 0.100 | 0.802 ± 0.090 |
| Supervised model (10 subjects) | 0.838 ± 0.029 | 0.911 ± 0.018 | 0.763 ± 0.068 | 0.860 ± 0.061 | 0.715 ± 0.074 | 0.825 ± 0.070 |
| Supervised model (20 subjects) | 0.838 ± 0.032 | 0.911 ± 0.020 | 0.798 ± 0.077 | 0.882 ± 0.067 | 0.771 ± 0.072 | 0.862 ± 0.066 |
| Supervised model (40 subjects) | 0.844 ± 0.031 | 0.914 ± 0.020 | 0.854 ± 0.084 | 0.916 ± 0.069 | 0.813 ± 0.069 | 0.889 ± 0.063 |
| TgtFreeHarmony (ours) | **0.839 ± 0.024** | **0.911 ± 0.015** | **0.822 ± 0.045** | **0.899 ± 0.030** | **0.805 ± 0.079** | **0.882 ± 0.070** |

**Table S2.** Segmentation performance comparison between our method (TgtFreeHarmony) and fine-tuned models trained with varying numbers of labeled subjects (1, 5, 10, and 20) from each source domain. Scores that surpass TgtFreeHarmony are underlined.

| Dataset | OASIS-3 | | | | | |
|---|---|---|---|---|---|---|
| | Sonata → Tim Trio | | Vision → Tim Trio | | Magnetom Vida → Tim Trio | |
| | IoU↑ | Dice↑ | IoU↑ | Dice↑ | IoU↑ | Dice↑ |
| Fine-tuned model (1 subjects) | 0.755 ± 0.074 | 0.852 ± 0.051 | 0.786 ± 0.070 | 0.874 ± 0.055 | 0.799 ± 0.084 | 0.880 ± 0.070 |
| Fine-tuned model (5 subjects) | 0.829 ± 0.039 | 0.905 ± 0.024 | 0.787 ± 0.078 | 0.875 ± 0.068 | 0.819 ± 0.090 | 0.892 ± 0.073 |
| Fine-tuned model (10 subjects) | 0.863 ± 0.028 | 0.925 ± 0.017 | 0.814 ± 0.079 | 0.893 ± 0.067 | 0.837 ± 0.073 | 0.903 ± 0.064 |
| Fine-tuned model (20 subjects) | 0.874 ± 0.028 | 0.932 ± 0.017 | 0.864 ± 0.077 | 0.923 ± 0.063 | 0.877 ± 0.086 | 0.927 ± 0.070 |
| TgtFreeHarmony (ours) | **0.839 ± 0.024** | **0.911 ± 0.015** | **0.822 ± 0.045** | **0.899 ± 0.030** | **0.805 ± 0.079** | **0.882 ± 0.070** |

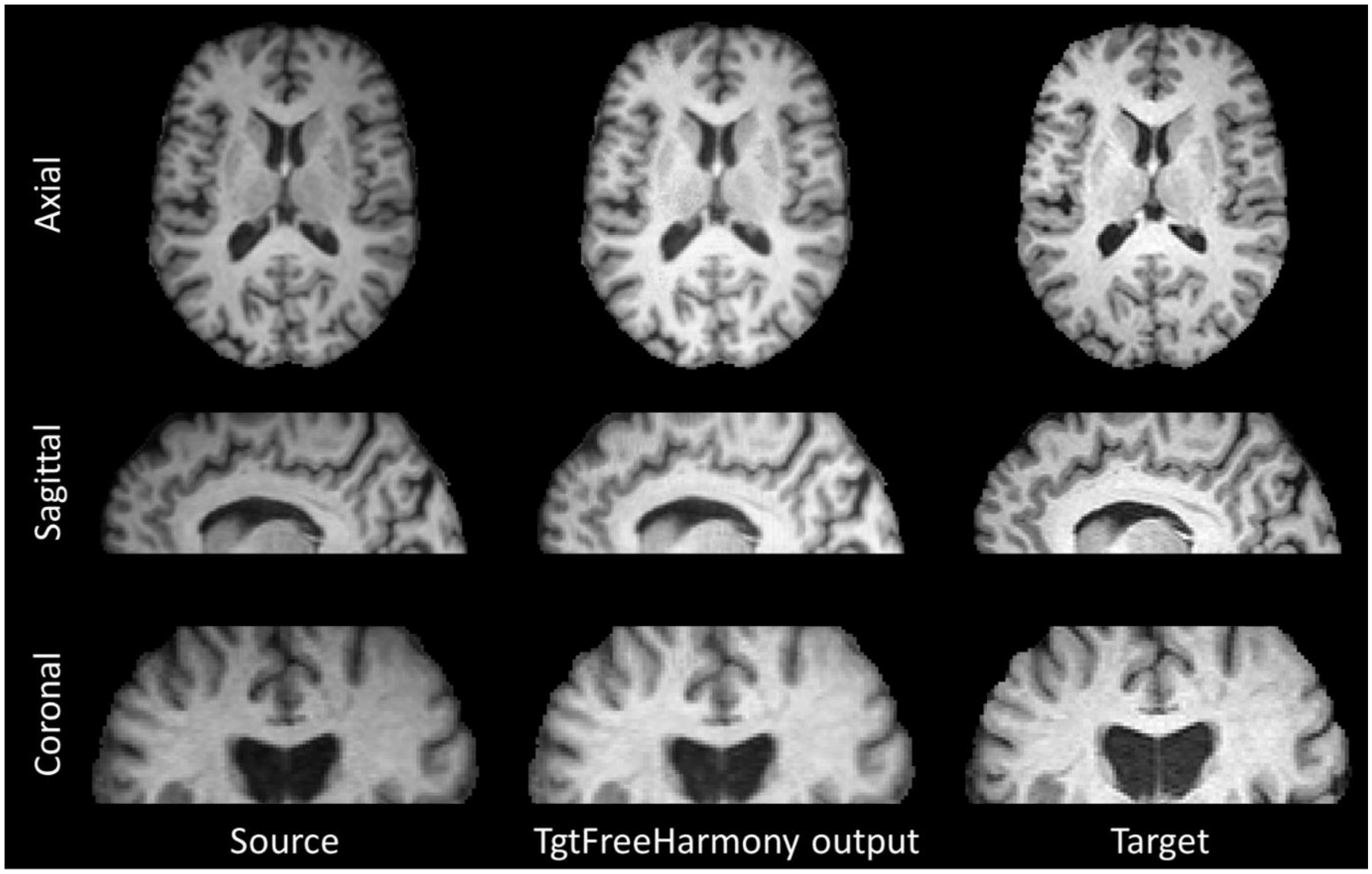

**Figure S4.** Harmonization output with the original source and target images is shown in the axial, sagittal, and coronal view. Although TgtFreeHarmony operates on 2D slices, applying the single estimated style vector uniformly across all slices yields a consistent slice appearance across the volume.

## References

Chen, X., Duan, Y., Houthooft, R., Schulman, J., Sutskever, I., Abbeel, P., 2016. Infogan: Interpretable representation learning by information maximizing generative adversarial nets. Advances in neural information processing systems 29.

Ronneberger, O., Fischer, P., Brox, T., 2015. U-net: Convolutional networks for biomedical image segmentation, Medical Image Computing and Computer-Assisted Intervention–MICCAI 2015: 18th International Conference, Munich, Germany, October 5-9, 2015, Proceedings, Part III 18. Springer, pp. 234–241.